\documentclass[aps,prl,twocolumn,superscriptaddress,showpacs,longbibliography,lengthcheck]{revtex4-1}

\usepackage{amsmath,amssymb}
\usepackage[dvipdfmx]{graphicx}

\usepackage{color} 

\usepackage{txfonts}
\usepackage{xspace}  
\usepackage{multirow}
\usepackage{dcolumn} 

\usepackage{xspace}  

\begin{document}

\title{Symmetries Created by Random Interactions\ \\
Ultimacy of ``More Is Different'' \, }


\newcommand{\ariken}{      \affiliation{RIKEN Nishina Center, 2-1 Hirosawa, Wako, Saitama 351-0198, Japan}}
\newcommand{\acns}{        \affiliation{Center for Nuclear Study, The University of Tokyo, 7-3-1 Hongo, Bunkyo, Tokyo 113-0033, Japan}}
\newcommand{\aut}{         \affiliation{Department of Physics, The University of Tokyo, 7-3-1 Hongo, Bunkyo, Tokyo 113-0033, Japan}}
\newcommand{\akul}{         \affiliation{KU Leuven, Instituut voor Kern- en Stralingsfysica, 3000 Leuven, Belgium}}

\newcommand{\aemt}{\email{Corresponding author: otsuka@phys.s.u-tokyo.ac.jp}}  
 
\author{Takaharu Otsuka}   \aemt \aut \ariken \akul 
\author{Noritaka Shimizu}   \acns 

\date{\today}

\begin{abstract}  
\bf{
The dominance (preponderance) of the 0$^+$ ground state for random interactions is shown to be a consequence of certain random interactions with chaotic features.    These random interactions, called chaotic random interactions, impart a symmetry property to the ground-state wave function: an isotropy under an appropriate transformation, such as zero angular momentum for rotation.  
Under this mechanism, the ground-state parity and isospin
can also be predicted in such a manner that positive parity is favored over 
negative parity and the isospin $T=0$ is favored over higher isospins.  
As chaotic random interaction is a limit with no particular dynamics at the level of two interacting particles, this realization of isotropic symmetry in the ground state can be considered as the ultimate case of many-body correlations.
A possible relation to the isotropy of the early universe is mentioned.
} 
\end{abstract}

\maketitle

A highly intriguing subject is the exploration of new aspects of the many-body 
system that do not directly follow from the interaction between constituent 
particles, which has been described as ``More Is Different'' by Anderson \cite{Anderson72}. 
We shall, in this Letter, examine one such subject in terms of the so-called random interactions, in a quest for the ultimacy of ``More Is Different.''

We consider systems of fermions interacting through a two-body interaction, $v$.  
Those fermions are protons and neutrons presently, but this is not an essential prerequisite in this work.  They are assumed to move in a set of single-particle orbits, $j_1, j_2, ..$.
A random interaction means that with $k_1, k_2, k_3$, and $k_4$ denoting the single-particle states of those orbits, two-body matrix elements 
$\langle k_1, k_2 | v | k_3, k_4 \rangle$ are given by random numbers, while   
certain constraints due to symmetry principles may apply.  These types of random interactions were shown by Bohigas {\it et al.} to lead to quantum chaos \cite{Bohigas84}.

Johnson, Bertsch, and Dean have shown a striking feature of random interactions
 \cite{JBD98}: 
a set of Configuration Interaction (CI;  {\it shell model} in nuclear physics) calculations were performed with various random interactions, and an ensemble of
ground-state spins was obtained.  One can expect that the fraction of the ground-state spin that is $J$=0 in this ensemble is, because of the randomness, close to the fraction of the $J$=0 subspace in the entire many-body Hilbert space for this system.  In an example considered in \cite{JBD98}, this fraction is 9.8 \%.  Contrary to this natural expectation, the value obtained by the CI calculation with random interactions, turned out to be 76 \%.   This huge difference was quite surprising, and has attracted much attention since then, inspiring many studies, for instance,  
\cite{BF00,MVZ00,KZC00,KPJ01,FI02,ZAY02,HVZ02,ZPBFA02,VZ02,Zrev04,Zhao04,Zhao04PRC,Shimizu07,Kirson07,Johnson12}.
This finding has been referred to as the $J$=0 preponderance, and its major features include the following: 
(i) it occurs, even if the pairing interaction is switched off, 
(ii) no collectivity can be seen, and 
(iii) no general connection to symmetries has been established.

Although the appearance of these features has been confirmed, the underlying common mechanism for the $J$=0 preponderance is not known, apart from special cases with some algebraic structures \cite{FI02,ZAY02}.
In fact, this is what the abstract of a review article states \cite{Zhao04}: ``a more fundamental understanding of the robustness of 0 g.s. dominance is still out of reach.''  
We shall present, in this Letter, a possible answer to this difficult question and its general consequences.


As in earlier works, the following Hamiltonian is taken in the present CI, or shell-model,  calculations:
\begin{equation}
H \, = \, \Sigma_{j_1 j_2 j_3 j_4} v^{(L)}_{j_1 j_2 j_3 j_4} 
     \Bigl( A^{\dagger (L)}(j_1, j_2)  \, A^{(L)} (j_3, j_4) \Bigr),
\label{eq:H}
\end{equation}
where $j_1, j_2, j_3$, and $ j_4$ denote single-particle orbits, and $v^{(L)}_{j_1 j_2 j_3 j_4}$ stands for the two-body matrix element (TBME) with the bra (ket) state comprising two particles in the orbits $j_1$ and $j_2$  ($j_3$ and $j_4$) coupled to the total angular momentum  $L$. 
Here, the operator $A^{\dagger (L)}$ creates a pair of fermions coupled to the angular momentum $L$, as shown by the brackets $[\,\,]$,  with a proper normalization, 
\begin{equation}
A^{\dagger (L)}(j_1, j_2) \, \propto \, [a^{\dagger}_{j_1} a^{\dagger}_{j_2} ]^{(L)},
       \label{eq:A}
\end{equation}
and $A^{(L)}$ is its conjugate.  Consequently, the TBME $v^{(L)}_{j_1 j_2 j_3 j_4}$ is properly normalized and antisymmetrized, too.     The outer parenthesis $\bigl( \, \bigr)$ in eq.~(\ref{eq:H}) denotes a scalar product.   
In eq.~(\ref{eq:H}), some mathematical details have been omitted for brevity; e.g.,  double counting is avoided in the summation.

In the discussions below, the $v^{(L)}_{j_1 j_2 j_3 j_4}$'s are  random numbers, unless otherwise specified.
These random numbers are in fact Gaussian-distribution-generated random numbers, and will be called a two-body random matrix ensemble (TBRE).
In the Hamiltonian above, single-particle energies are assumed to be
zero because their effects are not essential in this study.

In considering the properties of the ground state of such random interactions, we proceed with the following ansatz consisting of three parts.
\begin{enumerate}
\item
{\it Random mixing}:\\ 
  A certain class of random interactions mix various states strongly
  in a chaotic way without preferences, in the eigenstates of a many-body system.
  
  Some other random interactions can have, in general, particular dynamics 
  {\it accidentally}, as a consequence of random sampling.
  Such interactions are outside the scope of this ansatz, and are distinguished from 
  the random interactions with no (or negligible) dynamics.  For this purpose, the 
  latter, i.e., the random interactions without any specific dynamics, will be 
  called {\it chaotic random interactions} hereafter.
   
  We shall return to the classification of chaotic and non-chaotic random 
  interactions later.
 
\item
{\it Isotropy (Invariance)}: \\
The ground state $\Psi_g$ can be generated by the following operation on an 
arbitrary state, $\Psi_0$,  unless $\Psi_0$ is orthogonal to $\Psi_g$:
\begin{equation}
\Psi_g \, \propto \, \Bigl\{ \, \lim_{\beta \to \infty} {\mathrm exp}\big[ -\beta H\big] \, \Bigl\}  
 \,\, \Psi_0 .       \label{eq:R}
\end{equation}
This equation implies that multiple actions of $H$ bring the state on the right-hand side close to the ground state.    Different many-body state vectors are then mixed.    With the chaotic random interaction, this mixing occurs without any  particular dynamics.

We postulate that different orientations in some coordinates that describe the system (e.g., the ordinary three-dimensional coordinates) cannot be distinguished in the ground state at the limit of the ultimate chaotic mixings mentioned above.  
In other words, any orientation cannot have special significance because of random couplings with many state vectors without any specifications. 
Although this postulated feature may not arise perfectly in actual examples, 
it should be further enhanced, as a general trend, by increasing the number of single-particle orbits and/or increasing the number of active particles, as the mixings of different state vectors become more complex and stronger.       
If there are certain restrictions, e.g., due to the particle number, the ground-state wave function tends to be as close to be isotropic (or invariant) as possible.    
Finally we note that the postulated feature is absent in the other limit of only two particles.
  
This isotropy postulate appears very likely to be valid, and will be examined with some numerical simulations in this Letter.  

\item
{\it Quantum numbers:}\\
  The isotropy/invariance, if achieved, determines the quantum number(s) of the ground state, as we shall see soon.  
   
\end{enumerate}

We now investigate an application of this ansatz, 
the CI (shell-model) calculation shown in \cite{JBD98}: 
A six-neutron system was placed 
in the model space spanned by the $1d_{5/2,3/2}$ and $2s_{1/2}$ orbits, which is
nothing but the $sd$ shell.
The CI calculation was performed for random interactions, and produced 
a large probability, 76 \%, of the $J$=0 ground state \cite{JBD98}.  
The appropriate transformation is rotation in the usual three-dimensional configuration space including the spin of the neutron.
If the randomly sampled interaction happens to be a chaotic random interaction, 
it drives all orientations to be equal in the ground state, giving rise to an angular isotropic wave function.  
The angular isotropic wave function implies a $J$=0 state.
The feature described in  \cite{JBD98} is thus explained well by the present scheme, 
with a natural assumption that a chaotic random interaction is obtained more likely in the random sampling.
Note that the angular isotropic wave function is possible with an even number of neutrons, but not with an odd number of neutrons.  We shall comment on this point later.

 
\begin{figure}[bt]
  \centering
  \includegraphics[width=8.5cm]{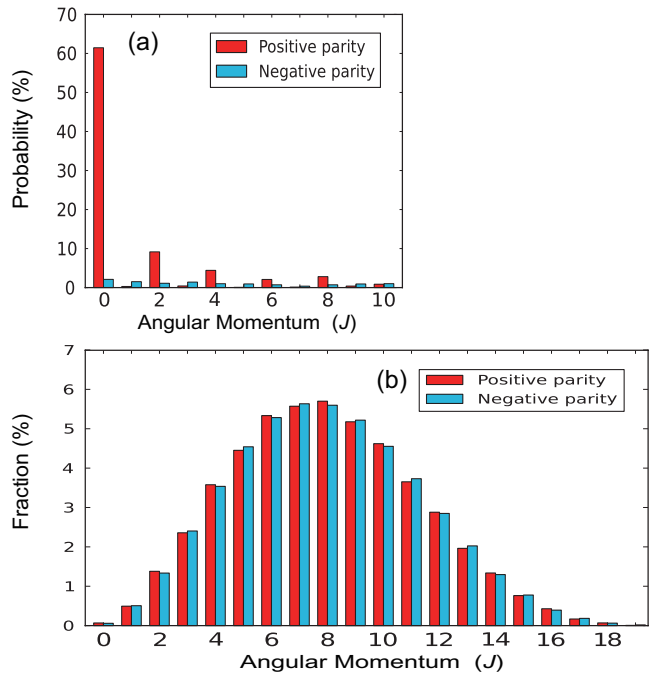}
  \caption{  
  (a) Probability distribution of the appearance of the
   ground state with given angular momentum, $J$, and parity, $\pi$, and 
   (b) fraction of the dimension of the subspace of each $J^{\pi}$ value  
   over the dimension of the whole Hilbert space,  
   for the system of ten neutrons in the model space comprising the 
   $1f_{5/2}$, $2p_{3/2,1/2}$, and $1g_{9/2}$ orbits. 
   The positive parity states are indicated by red histograms,
   and the negative parity states by blue ones.
  }   
  \label{fig:parity}  
\end{figure}  


We shall apply the same idea to a model space comprising 
orbits of positive and negative parities.  Up to now, we have not considered parity, but the present ansatz certainly covers the parity transformation.  
This transformation can be expressed by the simultaneous change
of three coordinates: (x, y, z) $\rightarrow$ (-x, -y, -z).  The chaotic random interaction tends to make amplitudes of the wave function for  (x, y, z) and (-x, -y, -z) 
equal.   
For a system with an even number of neutrons, 
the $J$=0 (i.e., isotropic) ground state can occur, and this property of equal amplitudes can arise.    
The preponderance of the $J^{\pi}$=0$^+$ ($\pi$:parity) ground state is thus expected.
Figure \ref{fig:parity}(a) shows the probability
distribution of the appearance of the ground state with given
angular momentum and parity, $J^{\pi}$.
The single-particle orbits are $1f_{5/2}$, $2p_{3/2,1/2}$, and $1g_{9/2}$,
and 10 neutrons are taken.  The ground state is predominantly of
$J^{\pi} = 0^+$.  
Figure \ref{fig:parity}(b) depicts  the fraction of the dimension of the subspace of each $J^{\pi}$ value  over the dimension of the whole Hilbert space of the present system.     
One finds a completely different pattern from Fig.~\ref{fig:parity}(a), with a huge bump around $J\sim$ 8.  Figure \ref{fig:parity}(b) shows no practical difference between the opposite parities.   A trend similar to Fig.~\ref{fig:parity}(a) is shown in \cite{Zhao04} without the present explanation.


\begin{figure}[bt]
  \centering
\includegraphics[width=7cm]{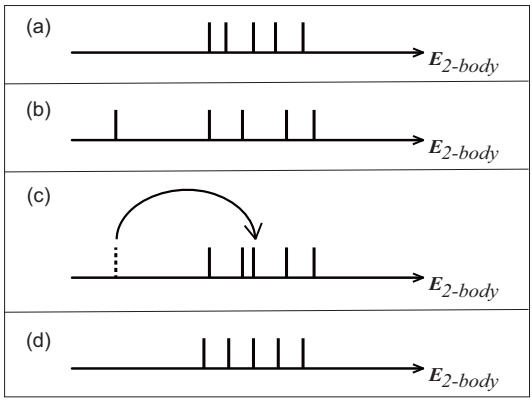}
  \caption{Schematic illustration of two-body spectra. 
  Vertical bars indicate energy eigenvalues for
  (a) all eigenvalues close to each other, (b) one eigenvalue far away, 
  (c) quantum mutant shifted to the right, and 
  (d) random interaction with confined equi-spaced eigenvalues.
  } 
\label{fig:2-body}
\end{figure}  

We now demonstrate how the chaotic random interaction works for the present preponderance problem.  
For this purpose, we introduce energy eigenvalues of the two-body system,
$E_{2-body}$.
Note that TBMEs are changed by a unitary transformation among basis vectors, but the eigenvalues do not change.       
The eigenvalues can be transformed back to TBMEs in eq.~(\ref{eq:H}) by using the wave functions of all two-body eigenstates.
We classify $E_{2-body}$
according to the $L^{\pi}$ of the two--body system (see eqs.~(\ref{eq:H}, \ref{eq:A})).
Figures \ref{fig:2-body}(a) and (b) schematically indicate two typical cases.
Panel (a) shows all $E_{2-body}$'s for some arbitrary $L^{\pi}$, which are rather close to each other.  
This belongs to the category of chaotic random interaction.  
We expect the $J^{\pi}$=0$^+$ preponderance for the ground state, based on the arguments of the above ansatz.
In panel (b), however, although most of the $E_{2-body}$'s are close to each other, the lowest one is far away from the rest.  In fact, in panel (b), the lowest eigenvalue
can produce a certain dynamical effect, resulting in a $J \neq 0$
ground state, if it occurs with $L \neq 0$.  

For the case of panel (b), we shift the lowest eigenvalue to the middle
of the rest (actually to $E_{2-body}$=0, but its precise value is not important), as shown in Fig.~\ref{fig:2-body}(c).  
The shift is only for this particular eigenvalue, and the two-body wave function is
kept unchanged.   Thus, the randomness is maintained.  Such eigenvalues far from the rest of the eigenvalues are called {\it quantum mutants} for the sake of clarity. 
Although the ground state may not be  $J$=0 originally, the $J$=0 ground state may appear after the present shift of a quantum mutant, as the interaction can change to a chaotic random interaction.


\begin{figure}[bt]
  \includegraphics[width=7.0cm]{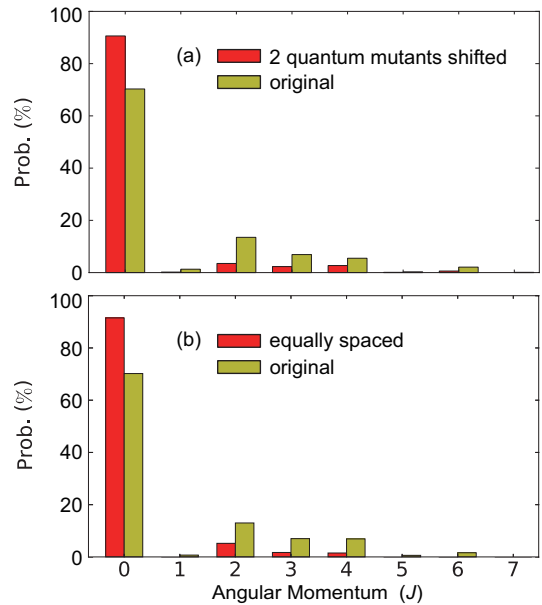}
  \caption{Probability distribution of the appearance of the ground state
    with a given angular momentum, $J$.  
    Olive green histograms are obtained from the original TBRE.
    Red histograms are obtained (a) from double quantum mutant shiftings  
     (see Fig.~\ref{fig:2-body}(c)), or 
    (b) from equally spaced $E_{2-body}$'s (see Fig.~\ref{fig:2-body}(d)).    
    }
\label{fig:mutant}
\end{figure}

To observe consequences of removing a quantum mutant, the CI calculations are  carried out for six neutrons in the $sd$ shell, the same setup as  in \cite{JBD98}.   
Figure~\ref{fig:mutant}(a) shows two types of calculations.  In one type   (olive green  histograms), the random interactions are provided by the original TBRE.   The results exhibit the $J$=0 preponderance, but one finds certain probabilities for $J \ne$0 ground states.

To investigate the possible origin of those $J$$\ne$0 ground states, we carry out the quantum mutant shifting shown in Fig.~\ref{fig:mutant}(c).  First, the eigenvalues of the two-body system are calculated, for a given random interaction, for all $L^{\pi}$ values except $L^{\pi}$=0$^+$.  For a specific $L^{\pi}$
where the number of the eigenstates is $n(L^{\pi})$, these eigenvalues are denoted by $E^{(2-body)}(L^{\pi};\,i\,=1, ..., n(L^{\pi}))$.  The most relevant quantum mutant is chosen in terms of 
\begin{eqnarray}
&E^{(2-body)}_{min} 
\,=\, & min\Bigl\{ E^{(2-body)}(L^{\pi}, \,i=1, ..., n(L^{\pi})) \,/ \sqrt{n(L^{\pi})} \,; \nonumber \\
& & \,\,\,\,\,\,\,\,\,\,\,\,\,  {\rm for} \,{\rm all} \, L^{\pi}\ne{\mathrm 0}^+ \Bigr\}.
\label{eq:min}
\end{eqnarray}
The quantum mutant thus identified is represented schematically by the vertical bar at the far left of Fig.~\ref{fig:2-body}(c), although it was taken from an assembly of different $L^{\pi}$ cases.   We then shift the eigenvalue of this quantum mutant, retaining its wave function.  We can calculate the TBMEs corresponding to the new eigenvalues, with which we perform a CI calculation to see the ground state.    
This process is applied to all cases with $J \ne$ 0 ground states of the above example: 10,000 random interactions are taken and $J \ne$ 0 ground states appear  in approximately 30\% of the cases.  We apply the quantum mutant shifting for the $J \ne$ 0 ground states, and approximately half of those cases show the change from $J\ne$0 to $J$=0 ground states.    
This is exactly what is expected from the change due to an interaction with some dynamics (given by random sampling) to a chaotic random interaction.  

We repeat the same process for the remaining 15\% cases (of the original 10,000 cases) with $J \ne$ 0 ground states.  
That is, the second quantum mutant is shifted in the same way.  As a result, 6\% cases out of the 15\% cases gain $J$=0 ground states.  Thus, after removing the two furthest  quantum mutants, a total of 91\% of the cases show $J$=0 ground states, implying a considerable enhancement of the $J$=0 preponderance.
 
The removal of the quantum mutants is one way of transforming non-chaotic random interactions to chaotic ones.  Figure~\ref{fig:mutant}(b) depicts another approach for the same system.
As with panel (a), in one set of the calculations, the random interactions are provided by the original TBRE (olive green histograms).  
In the other set, all $E_{2-body}$'s of a given $L^{\pi}$ are shifted such that they are equally spaced as shown in Fig.~\ref{fig:2-body}(d), with their mean value being zero and their variance being a fixed value common to all $L^{\pi}$'s.  
Of course, the wave functions of two-body systems are not changed from the original ones by the TBRE, maintaining the randomness.  
The TBMEs are calculated from those eigenvalues, resulting in 
 {\it random interactions with confined equi-spaced (two-body) eigenvalues.
These interactions 
carry essential features of chaotic random interactions, while the equal  spacing
is only for the simplicity.}
The obtained probability distribution, shown by red histograms in 
Fig. \ref{fig:mutant}(b),
demonstrates a pronounced $J$=0 preponderance:  
the $J$=0 probability is increased by $\sim$20$\%$ from that by the original TBRE interaction (see olive green histograms).  We also notice similarities between panels (a) and (b), which suggests that chaotic random interactions produce quite robust characteristics, irrespective of the details.  
  
\begin{figure}[bt]
  \centering
\includegraphics[width=7.5cm]{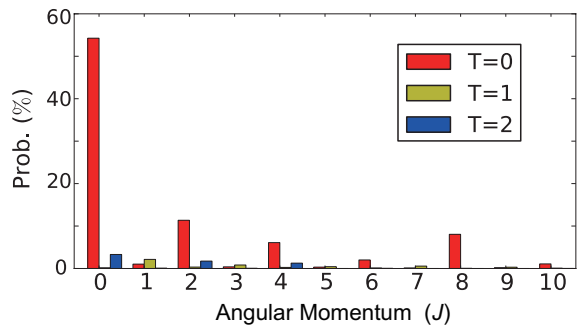}
\\
  \caption{Probability distribution of the appearance of the ground state
    with a given angular momentum, $J$ and isospin $T$.  
  } 
\label{fig:isospin}
\end{figure}  

The present ansatz can be applied to a more abstract symmetry such
as isospin, $T$.  The isospin stands for the symmetry between a proton and 
a neutron like 1/2-spin up and down, and the rotation in the isospin space
mixes a proton and a neutron.   
The wave function of a $T=0$ state remains unchanged by this rotation, implying an isotropy (or invariance) in the isospin space.  The chaotic random interaction favors ground states with this isotropy.
We then expect a preponderance of the $T=0$ ground state, if this is possible: an equal number of protons and neutrons in the same
single-particle space.   This isotropy can coexist with the isotropy of the three-dimensional configuration space,  
and therefore the preponderance of the $J$=$T$=0 ground state is predicted.   
Figure \ref{fig:isospin} shows the probability
distribution of the ground state with given
angular momentum, $J$, and isospin, $T$.
The single-particle space is the $sd$ shell, and four protons and
four neutrons are taken.
The ground state is in fact dominated by $T=0$ as predicted, 
even if $J \ne 0$.
This phenomenon was also discussed in \cite{HVZ02}.

If the number of protons is not equal to the number of neutrons,  $T$=0 states cannot be created, and the minimum isospin is half their difference.  If there are two more neutrons than protons, for instance, this difference disturbs the above isotropy mechanism, although states closer to the isotropy are favored, giving more probabilities to the states of $T$ lower.   This tendency has been reported, e.g., in \cite{Kirson07}.

We have discussed systems with an even number of protons and another even number of neutrons, including zero.  If either number is odd, 
$J$=0 is not possible, and the situation becomes more complex.
Furthermore, the present mechanism for the angular momentum and parity combined (see Fig.~\ref{fig:parity}) becomes inapplicable to such odd-number cases,   
providing almost equal probabilities for both parities (see a review in \cite{Zhao04}).

The present mechanism can be applied not only to the ground state
but also to the state highest in energy, as proved by changing the sign of
the interaction.
Concerning bosons, the present chaotic mixing is suppressed because of 
possible boson condensation \cite{BF00,KZC00}.

In summary, we propose in this Letter a mechanism for chaotic realization of isotropy (or invariance) in the ground-state wave function for various many-body systems.  
This is a simple idea that is applicable in many cases, covering the angular momentum, the parity, and the isospin.  This mechanism imparts a symmetry, i.e., isotropy, to the wave functions.        
The chaotic random interaction has no particular dynamical features but provides this symmetry property in many-body systems.  This intriguing relation between the force and the structure may be nicely described by the {\it Ultimacy of ``More is Different''} \cite{Anderson72}.  
The preponderance probability, e.g., of the $J$=0 ground state, is related somehow to the fraction of the chaotic random interactions in the interactions sampled randomly.    
Although we have a basic understanding of this, a precise separation between chaotic and non-chaotic random interactions may be rather complex, depending also on the system, and is an open problem that may require a new mathematical recipe. 
  
In regard to the angular momentum, the third type of the major origins of 0$^+$ ground state arises after
the pairing (or BCS) 0$^+$ state \cite{Bardeen1957,Bohr1958} and the 0$^+$ rotational-band head (Nambu-Goldstone mode)
\cite{bohr_mottelson_book2,Nambu1960,Goldstone1961,Goldstone1962}.  A search
for experimental examples is very intriguing, too, although the chaotic random interaction can represent an aspect of real nuclear forces.
The present idea can be generalized to other quantum many-body systems. 
Because of the overall picture on the ground state given by the present feature, its relation to the so-called uncertainty quantification is of much interest, see \cite{Yoshida18} as an example of CI studies.
Another intriguing conjecture is that the isotropy of the universe within 10$^{-5}$ deviation  of the Cosmic Microwave Background \cite{Smoot92} might be related to the chaotic  mixing in the universe at its very beginning as a quantum object.  This object must be cooled down to its ground state, and this could be achieved by transferring energies and other quantities to dark matter.  Of course, this is only a possibility, but it shows the wide relevance of the present work.   


%
\section{Acknowledgements}
T.O. is grateful for valuable comments from Prof. M. Honma and Dr. T. Abe.  
T.O. and N.S. acknowledge the support from MEXT in the form of ``Priority Issue on post-K computer" (Elucidation of the Fundamental Laws and Evolution of the Universe) (hp160211, hp170230, hp180179, hp190160) and JICFuS.
This work was supported in part by MEXT KAKENHI Grant No. JP19H05145, and in part by Grant-in-Aid for Specially Promoted Research (13002001).  
This work was supported in part by the RIKEN-CNS project on the nuclear structure 
calculation.

\end{document}